\documentclass[11pt,a4paper]{article}
\usepackage[utf8]{inputenc}
\usepackage{graphicx}
\usepackage{amsmath,amssymb,mathtools}
\usepackage{hyperref}
\hypersetup{colorlinks=true,linkcolor=blue,citecolor=blue,urlcolor=blue}
\usepackage{geometry}
\usepackage{multirow}
\usepackage{tikz}
\date{}

\begin{document}

\title{{\bf A two-mode model for black hole evaporation and information flow}}

\author{Erfan Bayenat$^{1}$\thanks{erfan.bayenat@iau.ac.ir}\,\, and\,\,\
Babak Vakili$^{2}$\thanks{ba.vakili@iau.ac.ir (corresponding author)} \,\,\,\\\\$^1${\small {\it Department of Physics, SR.C., Islamic Azad
University, Tehran, Iran}}\\$^2${\small {\it Department of Physics, CT.C., Islamic Azad
University, Tehran, Iran}} $^{}${}} \maketitle

\begin{abstract}
We develop and analyze a two--oscillator model for black hole evaporation in
which an effective geometric degree of freedom and a representative Hawking
radiation mode are described by coupled harmonic oscillators with opposite signs
in their free Hamiltonians. The normal--mode structure is obtained analytically
and the corresponding modal amplitudes determine the pattern of energy exchange
between the two sectors. To bridge the discrete and semiclassical pictures, we
introduce smooth envelope functions that provide a continuous effective
description along the geometric variable.
Numerical simulations in a truncated Fock space show that the two oscillators
exchange quanta in an approximately out--of--phase manner, consistent with an
effective conservation of $\langle n_x\rangle - \langle n_y\rangle$. The reduced
entropy $S_x(t)$ exhibits periodic growth, indicating entanglement generation.
These results demonstrate that even a minimal two--mode framework can capture
key qualitative features of energy transfer and information flow during
evaporation.
\vspace{5mm}\noindent
\end{abstract}

\section{Introduction}

Black holes occupy a central position in both classical and quantum gravity \cite{Ellis}.
Classically, they arise as exact solutions of Einstein’s field equations and
represent regions of spacetime from which no causal signal can escape. Their
geometric simplicity contrasts sharply with their rich physical behavior,
including the laws of black hole mechanics and their thermodynamic
interpretation~\cite{bardeen1973four, bekenstein1973black, bek2}. In particular,
the association of entropy with horizon area and temperature with surface
gravity indicates that black holes behave as thermodynamic systems, despite
being solutions of a purely classical field theory.

The quantum properties of black holes became dramatically clearer with
Hawking’s seminal discovery that quantum fields in curved spacetime predict
thermal radiation emitted from the horizon~\cite{hawking1975particle, Gary}. This
phenomenon, now known as Hawking radiation, implies that isolated black holes
are not truly stable but gradually evaporate over extremely long time scales.
The combination of Hawking radiation with black hole thermodynamics suggests
that the evaporation process is inherently quantum and encodes information
about the interaction between geometry and quantum fields. A key unresolved
issue in this context is the fate of information during evaporation and the
possible emergence of correlations or entanglement between the black hole
degrees of freedom and the emitted radiation. This question lies at the heart
of the modern black hole information problem \cite{kiefer2008instability, kiefer1999oscillator}.

A fully microscopic description of black hole evaporation requires a theory of
quantum gravity, which is not yet available in a complete form \cite{QG1, QG2}. Consequently,
various effective or phenomenological models have been developed to capture
qualitative aspects of black hole dynamics and information flow \cite{QBH1}-\cite{ob}. Among these,
a notable approach is the model introduced by Kiefer and collaborators,
who modeled the coupled system ``black hole + Hawking radiation'' in terms of
two interacting harmonic oscillators~\cite{kiefer2009blackhole}. In their
framework, one oscillator represents an effective geometric degree of freedom
associated with the black hole, while the second models a radiation mode. A
crucial ingredient of this construction is that the free Hamiltonians of the
two oscillators carry opposite signs, mimicking the idea that energy lost by
the black hole is gained by the radiative sector. This simple structure allows
one to study energy transfer, mode coupling, and entanglement generation in a
highly controlled quantum-mechanical setting. The model also provides a
testing ground for exploring information flow and decoherence mechanisms in
black hole evaporation without requiring a full quantum-gravity theory.

The present work develops and extends this two-oscillator phenomenological
framework in several directions. First, we derive the full normal-mode
solutions of the coupled system and analyze the role of the modal amplitudes
in governing the exchange of quanta between the geometric and radiative
sectors. We then introduce smooth envelope functions that interpolate the
discrete modal coefficients and enable a semi-continuous effective description
along a geometric variable. This construction bridges the gap between the
discrete Fock-space representation and a more geometrically motivated
effective dynamics. Next, we perform numerical simulations within a truncated
Fock space to compute occupation numbers, reduced density matrices, and the
time-dependent von Neumann entropy of the geometric subsystem. These
quantities allow us to characterize energy transfer, entanglement production,
and information flow during the evaporation-like process generated by the toy
model. The combined analytical and numerical results demonstrate that the
model captures several qualitative features expected from black hole
evaporation, including out-of-phase energy exchange and periodic growth of
entanglement.

\section{The model}

In order to capture the essential features of black hole evaporation in a simplified quantum framework, we follow the model approach of ~\cite{kiefer2009blackhole}. The key idea is to represent the coupled system ``black hole $+$ radiation'' by means of two interacting harmonic oscillators with opposite signs in their free Hamiltonians. This construction mimics the fact that the black hole degrees of freedom lose energy while the radiation field gains it. Their starting point is
the Schrödinger equation for an effective wave function 
$\Psi(m,\phi,t)$ depending on the black hole mass $m$ and a scalar
radiation degree of freedom $\phi$,

\begin{equation}
i\frac{\partial}{\partial t}\Psi(m,\phi,t)
= \hat{H}_{\rm BH}(m)\Psi + \hat{H}_{\rm rad}(\phi)\Psi
+ \hat{H}_{\rm int}(m,\phi)\Psi ,
\label{eq:schroedinger_mass_phi}
\end{equation}
where $\hat{H}_{\rm BH}$ is the effective Hamiltonian for the slowly varying
geometric degree of freedom, $\hat{H}_{\rm rad}$ describes the radiative
field mode responsible for Hawking emission and $\hat{H}_{\rm int}$ encodes
the interaction between geometry and matter.

Following \cite{kiefer2009blackhole}, we introduce two collective canonical variables, denoted $x$ and $y$, which serve as effective degrees of freedom for the evaporating black hole system. The variable $x$ is associated with the geometric degree of freedom (proportional to the black hole mass or radius in minisuperspace reductions), while $y$ encodes the radiation field mode responsible for Hawking emission. They obey standard canonical commutation relations

\begin{equation}
[x_i,p_j] = i\,\delta_{ij}, \qquad [x_i,x_j]=[p_i,p_j]=0.
\end{equation}
It can be shown that, after expanding the mass variable around a reference
value and introducing suitable canonical transformations, the system can be
approximated by two quadratic Hamiltonians, like those given in \cite{SF1}-\cite{SF3}. To leading order in this
approximation one obtains the effective Hamiltonian

\begin{equation}
H=\tfrac{1}{2}\left(p_x^2 + \omega_x^2 x^2\right) - \tfrac{1}{2}\left(p_y^2 + \omega_y^2 y^2\right) + g\,x y ,
\label{eq:Hxy}
\end{equation}
where $(x,p_x)$ represent the ``black hole oscillator'' and $(y,p_y)$ the ``radiation oscillator''. The relative minus sign ensures that energy lost from the $x$-sector appears as gained by the $y$-sector. One of the advantages of this representation is that the dynamical equations
derived from~\eqref{eq:Hxy} reproduce qualitative features known
from semiclassical analyses of black hole evaporation.
In particular, the oscillator variables $(x,y)$ define an effective
two–degree–of–freedom minisuperspace whose coupled dynamics accounts for energy exchange, mode mixing and entanglement generation
between the geometric and radiative subsystems.
Furthermore, expressing the Hamiltonian in normal-mode form provides a
compact description in terms of spectral coefficients $(A_j, B_j)$, which
encode the relative contributions of the two normal oscillators to the
physical modes $x$ and $y$.

In terms of annihilation and creation operators

\begin{equation}\label{ax}
a_x = \tfrac{1}{\sqrt{2}}(x+ip_x), \quad a_x^\dagger = \tfrac{1}{\sqrt{2}}(x-ip_x), \quad
a_y = \tfrac{1}{\sqrt{2}}(y+ip_y), \quad a_y^\dagger = \tfrac{1}{\sqrt{2}}(y-ip_y),
\end{equation}
the Hamiltonian \eqref{eq:Hxy} is equivalent to 

\begin{equation}\label{eq:H_general}
H=\omega_x\left(a_x^{\dagger}a_x+\tfrac12\right)-\omega_y\left(a_y^{\dagger}a_y+\tfrac12\right) + g\, xy \,.
\end{equation}
A useful feature of this Hamiltonian is that the operator

\begin{equation}
Q=a_x^\dagger a_x + a_y^\dagger a_y,
\end{equation}
is not conserved due to the coupling, but the difference

\begin{equation}
D = a_x^\dagger a_x - a_y^\dagger a_y,
\end{equation}
exhibits approximate constancy for weak coupling, which can be interpreted as a discrete analogue of total energy balance between geometry and radiation.

The Hilbert space of each oscillator is in principle infinite-dimensional. For numerical purposes, we introduce a cutoff $N_{\mathrm{cut}}$ and restrict the Fock space to basis states

\begin{equation}
\{|n_1,n_2\rangle : n_i = 0,1,\dots,N_{\mathrm{cut}}-1\}.
\end{equation}
This renders the Hamiltonian into a finite matrix of dimension $N_{\mathrm{cut}}^2$, suitable for exact diagonalization and time evolution on classical hardware. Later, this structure can be mapped to a quantum circuit simulation if desired. In order to model the onset of evaporation, we take as initial condition a product state of a coherent excitation in the black hole mode and the vacuum in the radiation mode

\begin{equation}
|\Psi(0)\rangle = |\alpha\rangle_1 \otimes |0\rangle_2 ,
\end{equation}
where $|\alpha\rangle$ is a coherent state defined by $a_x|\alpha\rangle = \alpha |\alpha\rangle$. The parameter $|\alpha|^2$ represents the average initial occupation of the black hole oscillator. The state at time $t$ is obtained via Schrödinger evolution

\begin{equation}
|\Psi(t)\rangle = e^{-iHt} |\Psi(0)\rangle,
\end{equation}
from which we may compute some observables such as occupation numbers: $\langle n_i(t) \rangle = \langle \Psi(t)| a_i^\dagger a_i |\Psi(t) \rangle$, quadratic variances: $\langle x_i^2(t) \rangle$ and $\langle p_i^2(t) \rangle$. Also, reduced density matrices $\rho_i(t)$ obtained by partial trace and the corresponding von Neumann entropies

\begin{equation}
S_i(t) = -\mathrm{Tr}\big( \rho_i(t) \ln \rho_i(t) \big).
\end{equation}
These entropies measure the degree of entanglement between the black hole and radiation oscillators. This formalism provides the basis for numerical experiments that capture information-flow and entropy production during the model evaporation.

\section{Dynamics and normal-mode analysis}
We now examine the classical and quantum dynamics of the quadratic Hamiltonian \eqref{eq:Hxy}. It is useful to first study the linearized classical equations of motion and identify the normal modes of the coupled system, the quantum evolution can then be built upon this structure via canonical quantization and Bogoliubov transformations. From Hamilton's equations applied to \eqref{eq:Hxy} we obtain

\begin{eqnarray}\label{F}
\left\{
\begin{array}{ll}
\dot x= p_x,\\
\dot p_x= -\omega_x^2 x - g y,\\
\dot y= -p_y, \\
\dot p_y= +\omega_y^2 y - g x.
\end{array}
\right.
\end{eqnarray}
The unusual sign in the $y$-momentum equation follows from the negative sign in the $y$-sector of the Hamiltonian. Introducing the phase-space vector $\xi=(x,p_x,y,p_y)^T$, the linear equations can be written compactly as

\begin{equation}
\dot \xi = J M \xi,
\end{equation}
where $J$ is the standard symplectic form

\begin{equation}
J=\begin{pmatrix}0 & 1 & 0 & 0\\ -1 & 0 & 0 & 0\\ 0 & 0 & 0 & 1\\ 0 & 0 & -1 & 0\end{pmatrix},
\end{equation}
and $M$ is the symmetric matrix of second derivatives (the Hessian of the quadratic form)

\begin{equation}
M=\begin{pmatrix} \omega_x^2 & 0 & g & 0\\ 0 & 1 & 0 & 0\\ g & 0 & -\omega_y^2 & 0\\ 0 & 0 & 0 & -1 \end{pmatrix}.
\end{equation}
Normal-mode frequencies $\Omega$ are found by seeking solutions of the form $\xi(t)\propto e^{i\Omega t}$. This leads to the secular equation

\begin{equation}
\det\big(V-\Omega^2 K\big)=0,
\end{equation}
where we have separated the Hamiltonian into kinetic and potential blocks with

\begin{equation}
K=\mathrm{diag}(1,-1), \qquad V=\begin{pmatrix}\omega_x^2 & g\\ g & -\omega_y^2\end{pmatrix}.
\end{equation}
Carrying out the determinant yields a quadratic equation for $\Omega^2$:

\begin{equation}\label{eq:Omega_secular}
(\Omega^2-\omega_x^2)(\Omega^2-\omega_y^2) + g^2 = 0.
\end{equation}
Thus

\begin{equation}
\Omega^2 = \frac{\omega_x^2+\omega_y^2}{2} \pm \frac{1}{2}\sqrt{(\omega_x^2-\omega_y^2)^2 - 4g^2}.
\end{equation}
The discriminant
\begin{equation}
\Delta = (\omega_x^2-\omega_y^2)^2 - 4g^2,
\end{equation}
controls the character of the solutions: for $\Delta\ge0$ the squared frequencies are real and the dynamics is oscillatory showing the stable normal modes, while for $\Delta<0$ one obtains complex frequencies and an instability exponential growth, signalling physically the runaway behaviour associated with the indefinite kinetic term when coupling is sufficiently strong.

If the normal-mode squared frequencies are positive, one can define canonical normal coordinates $(Q_1,P_1,Q_2,P_2)$ via a linear symplectic transformation

\begin{equation}
\xi = S \, \Xi, \qquad \Xi=(Q_1,P_1,Q_2,P_2)^T,
\end{equation}
with $S$ satisfying $S^T J S = J$. In these coordinates the Hamiltonian separates into two (possibly indefinite-sign) quadratic terms

\begin{equation}
H = \tfrac12 (P_1^2 + \Omega_1^2 Q_1^2) - \tfrac12 (P_2^2 + \Omega_2^2 Q_2^2 ).
\end{equation}
Canonical quantization promotes $(Q_i,P_i)$ to operators with $[Q_i,P_j]=i\delta_{ij}$, and the corresponding normal-mode annihilation operators are

\begin{equation}
b_j = \tfrac{1}{\sqrt{2}}(Q_j + i P_j), \qquad [b_j,b_k^\dagger]=\delta_{jk}.
\end{equation}
The relation between the original mode operators $(a_x,a_y)$ and the normal-mode operators $(b_1,b_2)$ is in general a Bogoliubov linear transformation mixing creation and annihilation parts when the transformation is not orthogonal in the complex sense. This mixing is the source of squeezing and particle production.

For Gaussian initial states (coherent or vacuum), the full quantum state remains Gaussian under quadratic evolution. The state is fully characterised by the first moments $\langle \xi\rangle$ and the covariance matrix

\begin{equation}
\sigma_{ij}(t) = \tfrac{1}{2}\langle \{\xi_i(t)-\langle\xi_i(t)\rangle,\xi_j(t)-\langle\xi_j(t)\rangle\}\rangle.
\end{equation}
The covariance matrix evolves under the symplectic flow as

\begin{equation}\label{sigma}
\sigma(t) = S(t)\,\sigma(0)\,S(t)^T,
\end{equation}
where $S(t)=\exp(J M t)$, is the symplectic propagator. Observables such as occupations and quadrature variances are directly obtained from entries of $\sigma(t)$. For a two-mode Gaussian state, the entanglement between modes can be characterised via the symplectic eigenvalues of the reduced covariance matrix. Let $\nu\ge\tfrac12$ denote the symplectic eigenvalue of the single-mode covariance matrix, the von Neumann entropy of a single mode (in natural units) is then

\begin{equation}\label{eq:gaussian_entropy}
S(\nu) = \left(\nu+\tfrac12\right)\ln\left(\nu+\tfrac12\right) - \left(\nu-\tfrac12\right)\ln\left(\nu-\tfrac12\right).
\end{equation}
Other measures, such as logarithmic negativity, can be expressed similarly in terms of symplectic invariants and provide alternative quantifications of entanglement generation during the evaporation process.

An alternative but equivalent formulation of the time evolution is obtained
in the Heisenberg picture. The operators satisfy

\begin{equation}
\frac{d}{dt} O(t) = i[H,O(t)] .
\end{equation}
Applying this to the canonical operators $(x,p_x)$ and $(y,p_y)$ gives the following coupled system of linear differential equations

\begin{equation}
\frac{d}{dt}
\begin{pmatrix}
x \\ p_x \\ y \\ p_y
\end{pmatrix}
=
\begin{pmatrix}
0 & 1 & 0 & 0 \\
-\omega_x^2 & 0 & -g & 0 \\
0 & 0 & 0 & -1 \\
-g & 0 & \omega_y^2 & 0
\end{pmatrix}
\begin{pmatrix}
x \\ p_x \\ y \\ p_y
\end{pmatrix}.
\end{equation}
The eigenvalues of the above matrix determine the normal-mode frequencies of the
coupled oscillators. Depending on the sign and magnitude of $g$, the
spectrum may exhibit instabilities which correspond, in the black hole
interpretation, to amplification of radiation at the expense of the
geometric degree of freedom. In terms of ladder operators, we obtain

\begin{eqnarray}
\left\{
\begin{array}{ll}
\dot{a}_x= -i \omega_x a_x - i g \tfrac{y}{\sqrt{2}}, \\\\
\dot{a}_y= +i \omega_y a_y - i g \tfrac{x}{\sqrt{2}}.
\end{array}
\right.
\end{eqnarray}
These relations allow one to compute correlation functions and entanglement dynamics directly in the operator formalism. The above analysis clarifies the mathematical structure of the dynamics and provides practical formulae for implementing and interpreting the numerical simulations described in the next section.

\section{Closed-form solution of the linear dynamics}

Starting from the Heisenberg/classical form linear system 
\begin{eqnarray} \label{eq:xx}
\left\{
\begin{array}{ll}
\ddot x + \omega_x^2 x + g\, y= 0,\\\\
\ddot y + \omega_y^2 y - g\, x= 0,
\end{array}
\right.
\end{eqnarray}
we seek separable solutions of the form $x(t)=X e^{i\Omega t}$, $y(t)=Y e^{i\Omega t}$. Inserting into
\eqref{eq:xx} yields the algebraic system

\begin{equation}\label{eq:eig_sys}
\begin{pmatrix}
-\Omega^2 + \omega_x^2 & g \\
-g & -\Omega^2 + \omega_y^2
\end{pmatrix}
\begin{pmatrix} X \\ Y \end{pmatrix}
= 0.
\end{equation}
Nontrivial solutions exist when the secular equation holds, which is nothing but the Eq.~\eqref{eq:Omega_secular}. Denote the two, generically distinct, roots for \(\Omega^2\) by \(\Omega_1^2\) and \(\Omega_2^2\), for each \(j=1,2\), an associated eigenvector \((X_j,Y_j)^T\) obeys

\begin{eqnarray}\label{F}
\left\{
\begin{array}{ll}
(\Omega_j^2-\omega_x^2) X_j = g\, Y_j, \label{eq:rel1}\\\\
(\Omega_j^2-\omega_y^2) Y_j= -g\, X_j. \label{eq:rel2}
\end{array}
\right.
\end{eqnarray}
We may choose a convenient normalization for the eigenvectors, for example set \(X_j=1\) and solve for \(Y_j\) (except when \(X_j=0\), then choose \(Y_j=1\)). In the stable case for which \(\Omega_j^2>0\), so \(\Omega_j\) real, the general real solution can be written as a superposition of the two normal modes

\begin{align}
x(t) &= \sum_{j=1}^2 \Big( A_j \cos(\Omega_j t) + B_j \sin(\Omega_j t)\Big) X_j, \label{eq:x_sol}\\
y(t) &= \sum_{j=1}^2 \Big( A_j \cos(\Omega_j t) + B_j \sin(\Omega_j t)\Big) Y_j. \label{eq:y_sol}
\end{align}
The constants \(A_j,B_j\) are fixed by initial data \(x(0),\dot x(0),y(0),\dot y(0)\). Therefore, the projection of initial conditions onto eigenvectors gives the linear system

\begin{equation}\label{eq:proj_init}
\begin{pmatrix}
X_1 & X_2 & 0 & 0 \\
0 & 0 & X_1 & X_2 \\
Y_1 & Y_2 & 0 & 0 \\
0 & 0 & Y_1 & Y_2
\end{pmatrix}
\begin{pmatrix} A_1 \\ A_2 \\ B_1 \\ B_2 \end{pmatrix}
=
\begin{pmatrix} x(0) \\ \dot x(0) \\ y(0) \\ \dot y(0) \end{pmatrix},
\end{equation}
which, provided the eigenvectors are linearly independent, can be inverted to obtain \(A_j,B_j\). A more explicit inversion is obtained by using orthogonality of the normal modes with respect to the appropriate bilinear form or by solving the \(4\times4\) linear system directly. To get the compact expression, let \(V\) be the $2\times2$ invertible matrix whose columns are the eigenvectors \((X_j,Y_j)^T\):

\begin{equation}
V = \begin{pmatrix} X_1 & X_2 \\ Y_1 & Y_2 \end{pmatrix},
\end{equation}
and define the modal coordinates \(q_j(t)=A_j\cos\Omega_j t+B_j\sin\Omega_j t\). Then

\begin{equation}
\begin{pmatrix} x(t) \\ y(t) \end{pmatrix} = V \begin{pmatrix} q_1(t) \\ q_2(t) \end{pmatrix}.
\end{equation}
Evaluating at \(t=0\) gives the linear relations

\begin{equation}\label{eq:init_pos}
\begin{pmatrix} x(0) \\ y(0) \end{pmatrix} = V \begin{pmatrix} A_1 \\ A_2 \end{pmatrix},
\end{equation}
and differentiating and evaluating at \(t=0\) yields

\begin{equation}\label{eq:init_vel}
\begin{pmatrix} \dot x(0) \\ \dot y(0) \end{pmatrix} = V \begin{pmatrix} \Omega_1 B_1 \\ \Omega_2 B_2 \end{pmatrix}
= V \,\Omega\, \begin{pmatrix} B_1 \\ B_2 \end{pmatrix},
\end{equation}
where \(\Omega=\mathrm{diag}(\Omega_1,\Omega_2)\). Therefore the coefficients \(A_j\) and \(B_j\) follow directly

\begin{align}
\begin{pmatrix} A_1 \\ A_2 \end{pmatrix}
&= V^{-1} \begin{pmatrix} x(0) \\ y(0) \end{pmatrix}, \label{eq:A_vec} \\
\begin{pmatrix} B_1 \\ B_2 \end{pmatrix}
&= \Omega^{-1} V^{-1} \begin{pmatrix} \dot x(0) \\ \dot y(0) \end{pmatrix}. \label{eq:B_vec}
\end{align}
Using the closed-form inverse of a $2\times2$ matrix, \(V^{-1} = \frac{1}{\Delta_V}\begin{pmatrix} Y_2 & -X_2\\ -Y_1 & X_1 \end{pmatrix}\), we obtain explicit scalar formulas

\begin{align}
A_1 &= \frac{Y_2\,x(0) - X_2\,y(0)}{\Delta_V}, \qquad
A_2 = \frac{-Y_1\,x(0) + X_1\,y(0)}{\Delta_V}, \label{eq:A_explicit} \\
B_1 &= \frac{1}{\Omega_1}\,\frac{Y_2\,\dot x(0) - X_2\,\dot y(0)}{\Delta_V}, \qquad
B_2 = \frac{1}{\Omega_2}\,\frac{-Y_1\,\dot x(0) + X_1\,\dot y(0)}{\Delta_V}, \label{eq:B_explicit}
\end{align}
where $\Delta_V = \det V = X_1 Y_2 - X_2 Y_1 \neq 0$ is the determinant of $V$. For practical calculations one may pick a normalization for eigenvectors to simplify \(\Delta_V\). As mentioned before, a common choice is to set \(X_j=1\) (if nonzero) and solve \(Y_j\) from \eqref{eq:rel1}, the formulas above then reduce to rational expressions of \(\omega_x,\omega_y,g\) and the initial data. For the unstable case which one of the \(\Omega_j^2\), is negative and the square root in the secular solution is imaginary, set \(\Lambda_j^2=-\Omega_j^2>0\) for those modes, the trigonometric functions for that mode are replaced by hyperbolic functions

\begin{equation}
q_j(t) = \tilde A_j \cosh(\Lambda_j t) + \tilde B_j \sinh(\Lambda_j t).
\end{equation}
This corresponds to exponential growth/decay of the corresponding normal coordinate and signals the dynamical instability discussed earlier.

Note that if \(\Delta_V=0\), the two eigenvectors are linearly dependent (degeneracy) and the modal decomposition above must be modified; typically one then needs to construct a generalized eigenvector (Jordan block treatment) or choose an alternative basis. In practice this situation corresponds to parameter choices where the secular matrix is defective.
On the other hand, if one or both \(\Omega_j^2\) are negative (unstable modes), replace \(\Omega_j\) by \(i\Lambda_j\) with \(\Lambda_j>0\), and the trigonometric functions in the modal solution by hyperbolic functions. The formulas \eqref{eq:A_explicit}--\eqref{eq:B_explicit} for \(A_j\) still apply for the cosine/hyperbolic-cosine coefficients, while the expressions for the coefficients multiplying the growing/decaying parts follow from the analogous inversion of the initial-value linear system.

Now, let us consider the quantum operator solution. The classical modal decomposition lifts directly to quantum operators by promoting initial amplitudes to operators. Equivalently, one can work with annihilation/creation operators: assemble the vector

\begin{equation}
\mathbf{A}(t) = \begin{pmatrix} a_x(t) \\ a_y(t) \\ a_x^\dagger(t) \\ a_y^\dagger(t) \end{pmatrix},
\end{equation}
which evolves linearly under the quadratic Hamiltonian

\begin{equation}\label{eq:bog_evol}
\mathbf{A}(t) = \mathcal{S}(t)\, \mathbf{A}(0),
\end{equation}
where \(\mathcal{S}(t)\) is a \(4\times4\) symplectic (Bogoliubov) matrix of the block form
\begin{equation}
\mathcal{S}(t) = \begin{pmatrix} U(t) & V(t) \\ V^*(t) & U^*(t) \end{pmatrix},
\end{equation}
with \(U(t)\) and \(V(t)\) being \(2\times2\) matrices satisfying the canonical commutation-preserving constraints \(U U^\dagger - V V^\dagger = \mathbb{I}\), \(U V^T = V U^T\). The entries of \(U\) and \(V\) can be obtained by projecting the modal solutions onto the ladder-operator basis. Explicitly, once \(A_j,B_j\) (or the modal time-dependence \(q_j(t)\)) are known, substitution into the relations $a_x(t) = \tfrac{1}{\sqrt{2}}(x(t) + i p_x(t))$ and $a_y(t) = \tfrac{1}{\sqrt{2}}(y(t) + i p_y(t))$, gives linear relations of the form \(\mathbf{A}(t)=\mathcal S(t)\mathbf{A}(0)\) (Eq.~\eqref{eq:bog_evol}), and hence explicit expressions for the Bogoliubov blocks \(U(t),V(t)\) in terms of \(\{X_j,Y_j,\Omega_j\}\) and the chosen normalization. Nonzero \(V(t)\) encodes mixing between creation and annihilation operators and is the hallmark of squeezing and particle (quanta) production in the quantum evolution. For Gaussian initial states (e.g. coherent state in mode \(x\) and vacuum in mode \(y\)), the time-dependent covariance matrix follows directly from \(\mathcal{S}(t)\) and the initial covariance; entropies and entanglement measures are then computable in closed form via the symplectic eigenvalues of the reduced covariance matrices (cf.\ Eq.~\eqref{eq:gaussian_entropy}). The closed-form modal solution given above is the basis for both analytical estimates (e.g.\ short-time squeezing rates, instability growth exponents) and for building numerically-stable expressions to compare with finite-cutoff simulations.

The covariance matrix evolves as $\sigma(t)=S(t)\sigma(0)S^T(t)$ with the
symplectic propagator $S(t)$ built from the modal cos/sin blocks. Thus each
element of the reduced covariance $\sigma_x(t)$ (a $2\times2$ matrix) admits
a closed-form expression in terms of $\{X_j,Y_j,\Omega_j,A_j,B_j\}$ and the
initial vacuum variances. Denoting

\begin{equation}\label{A1}
\sigma_x(t)=\begin{pmatrix}\sigma_{xx}(t)&\sigma_{xp}(t)\\\sigma_{xp}(t)&\sigma_{pp}(t)\end{pmatrix},
\end{equation}
one has symbolically

\begin{equation}
\sigma_{xx}(t)=\tfrac12 + \sum_{j,k} C^{(xx)}_{jk}\cos(\Omega_j t)\cos(\Omega_k t)
+ \sum_{j,k} D^{(xx)}_{jk}\sin(\Omega_j t)\sin(\Omega_k t),
\end{equation}
with analogous closed sums for $\sigma_{pp}$ and $\sigma_{xp}$. The symplectic
eigenvalue entering the von Neumann entropy is then

\begin{equation}\label{A2}
\nu_x(t)=\sqrt{\det\sigma_x(t)}.
\end{equation}
Hence $S_x(t)$ is an explicit (albeit somewhat lengthy) analytic function of time
built from trigonometric combinations of $\Omega_\pm t$.

For a Gaussian state in which the covariance contribution is the vacuum value
$1/2$ (or when first-moment contributions dominate, as in a large coherent state),
the occupation of the $x$-mode is expressible in terms of the first moments

\begin{equation}\label{B1}
\langle n_x(t)\rangle = \frac{1}{2}\big(\langle x(t)\rangle^2 + \langle p_x(t)\rangle^2\big).
\end{equation}
Inserting the modal expansions $x(t)=\sum_j q_j(t)X_j$ and $p_x(t)=\sum_j \dot q_j(t)X_j$, yields the closed finite-sum

\begin{equation}
\begin{split}
\langle n_x(t)\rangle
&= \tfrac{1}{2}\sum_{j,k=1}^2 X_j X_k
\Big[ A_j A_k \cos\Omega_j t\cos\Omega_k t + A_j B_k \cos\Omega_j t\sin\Omega_k t \\
&\quad + B_j A_k \sin\Omega_j t\cos\Omega_k t + B_j B_k \sin\Omega_j t\sin\Omega_k t \\
&\quad + \Omega_j\Omega_k\big(-A_j A_k \sin\Omega_j t\sin\Omega_k t
+ A_j B_k \sin\Omega_j t\cos\Omega_k t \\
&\qquad + B_j A_k \cos\Omega_j t\sin\Omega_k t - B_j B_k \cos\Omega_j t\cos\Omega_k t\big)
\Big],
\end{split}
\label{eq:nx_full}
\end{equation}
and similarly for $\langle n_y(t)\rangle$ (replace $X_j\to Y_j$ and $A_j\to$ appropriate modal projections). Upon collecting like terms, the above relation can be rearranged as a superposition of cosines with summed and differenced arguments

\begin{equation}
\langle n_x(t)\rangle = C_0 + \sum_{j\le k} \Big[C_{jk}^+ \cos((\Omega_j+\Omega_k)t) + C_{jk}^- \cos((\Omega_j-\Omega_k)t)\Big],
\end{equation}
where the coefficients $C_0,C_{jk}^\pm$ are algebraic functions of
$\{X_j,Y_j,\Omega_j,A_j,B_j\}$ (explicit forms follow from trigonometric product-to-sum identities
applied to Eq.~\eqref{eq:nx_full}).

For weak coupling and short times one may expand the dynamics perturbatively in
$g$ (or equivalently expand in $t$). Up to second order in $t$ the leading
behaviour is
\begin{align}
\langle n_y(t)\rangle &= \kappa_1\, (g t)^2 + \mathcal{O}(t^4),\\
S_x(t) &= \kappa_2\, (g t)^2 + \mathcal{O}(t^4),\label{B2}
\end{align}
where the coefficients $\kappa_{1,2}$ are expressible in closed form from the
initial moments of the $x$-mode (e.g.\ $x(0),p_x(0)$) and the bare frequencies
$\omega_x,\omega_y$. In particular, the quadratic growth of both $\langle n_y\rangle$
and $S_x$ reflects the fact that the coupling $gxy$ induces transitions at
second order in time (first nonvanishing population transfer appears at ${\cal O}(t^2)$).
This short-time result is useful for (i) predicting the initial slope of the
numerical curves and (ii) validating numerical implementations.

${\bullet}$ {\it{Numerical examples}}:
To end this section let us examine the formalism described above by some numerical examples to compute the explicit modal coefficients. As the first example we choose the parameter values
\(\omega_x=1.0\), \(\omega_y=0.8\) and \(g=0.15\). For an initial coherent
displacement \(\alpha=1\) the classical initial conditions (mapping
\(\alpha\mapsto x(0)=\sqrt{2}\,\alpha\)) are

\begin{equation}
x(0)=\sqrt{2}\approx1.41421,\qquad \dot x(0)=0,\qquad y(0)=0,\qquad \dot y(0)=0.
\end{equation}
Solving the secular equation \((\Omega^2-\omega_x^2)(\Omega^2-\omega_y^2)+g^2=0\)
yields two normal-mode squared frequencies \(\Omega_1^2\) and \(\Omega_2^2\).
Choosing the normalization \(X_j=1\) for each eigenvector, the corresponding
quantities are computed as in table \ref{tab:modal_example}.

\begin{table}[ht]
\centering
\begin{tabular}{rrrrrrr}
\hline
mode & $\Omega^2$ & $\Omega$ & $X_j$ & $Y_j$ & $A_j$ & $B_j$ \\
\hline
1 & 0.919499 & 0.958905 & 1.0 & -0.536675 & 1.986311 & 0.0 \\
2 & 0.720501 & 0.848823 & 1.0 & -1.863325 & -0.572098 & 0.0 \\
\hline
\end{tabular}
\caption{Modal parameters and initial modal coefficients for
\(\omega_x=1.0,\ \omega_y=0.8,\ g=0.15\) and initial data
\(x(0)=\sqrt{2},\ \dot x(0)=0,\ y(0)=0,\ \dot y(0)=0\).}
\label{tab:modal_example}
\end{table}
The determinant of the modal matrix is $\det V\approx -1.32665$, hence $V$
is invertible and the modal expansion is well-defined. The nonzero values
of $A_j$ (with $B_j=0$) reflect the fact that the initial displacement
has nonzero projection onto the cosine parts of the two normal modes,
while zero initial velocities give vanishing sine coefficients.

To illustrate the role of initial velocities (which produce nonzero sine-mode coefficients $B_j$), let us consider a numerical example with nonzero initial velocity.
We choose $\omega_x=1.0$, $\omega_y=0.8$, $g=0.15$ and initial data

\begin{equation}
x(0)=\sqrt{2}\approx1.41421,\qquad \dot x(0)=0.5,\qquad y(0)=0,\qquad \dot y(0)=0.
\end{equation}
Solving the secular equation and projecting the initial data onto the modal basis (with the
convenient normalization $X_j=1$) yields the modal parameters as shown in table \ref{tab:modal_example_velocity}.

\begin{table}[ht]
\centering
\begin{tabular}{rrrrrrr}
\hline
mode & $\Omega^2$ & $\Omega$ & $X_j$ & $Y_j$ & $A_j$ & $B_j$ \\
\hline
1 & 0.919499 & 0.958905 & 1.0 & -0.536675 & 1.986311 & 0.732364 \\
2 & 0.720501 & 0.848823 & 1.0 & -1.863325 & -0.572098 & -0.238291 \\
\hline
\end{tabular}
\caption{Modal decomposition for $\omega_x=1.0$, $\omega_y=0.8$, $g=0.15$ and
initial data with $\dot x(0)=0.5$.}
\label{tab:modal_example_velocity}
\end{table}
The determinant of the modal matrix is $\det V\approx -1.32665$, confirming invertibility.
The nonzero $B_j$ entries reflect the projection of the nonzero initial velocity onto the
modal sine components; these coefficients in turn contribute to time-dependent mixing
between creation and annihilation operators and lead to nontrivial squeezing dynamics.

\begin{figure}[ht]
\centering
\includegraphics[width=0.5\textwidth]{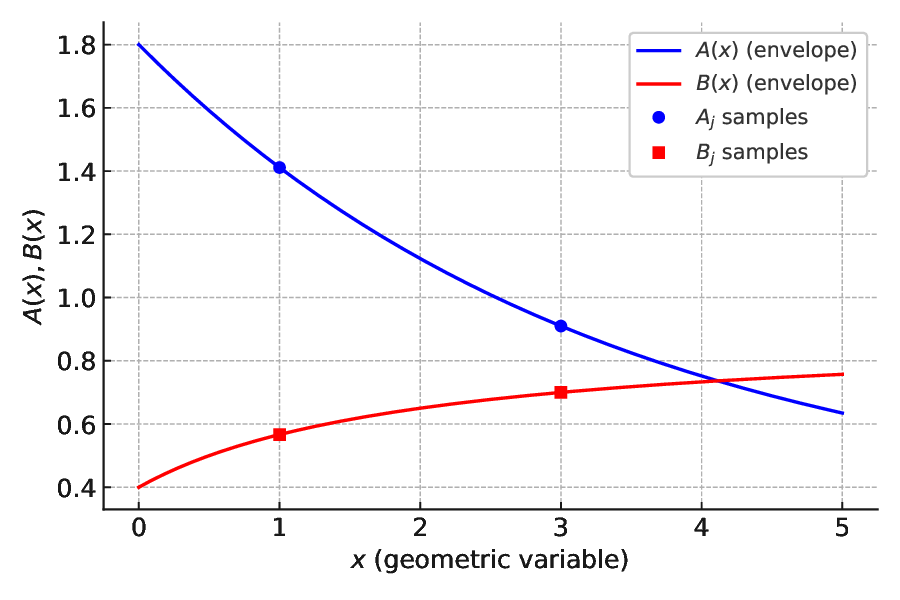}
\caption{
Continuous envelope functions $A(x)$ (blue) and $B(x)$ (red) constructed
from the discrete modal coefficients $A_j,B_j$.
The dots and squares correspond to the numerically obtained values
of the amplitudes for two representative modes, while the smooth curves
illustrate effective interpolations that describe how the black hole
and radiation sectors vary along the geometric variable~$x$.
These continuous profiles are used in the subsequent analytical
and numerical analysis to connect the discrete two-mode model
to a semiclassical, quasi-continuous description of the evaporation process.
}
\label{fig:A_B_envelope}
\end{figure}

${\bullet}$ {\it{Geometric interpretation}}: In order to relate the discrete two-mode dynamics to an effective
semiclassical description, we introduce continuous envelope functions
$A(x)$ and $B(x)$ representing smooth extensions of the modal coefficients
$A_j$ and $B_j$.
Although only a few discrete amplitudes are available from the numerical
solutions (corresponding to the dominant and subdominant normal modes),
their values capture the essential trend of the coupled oscillators.
Hence, $A(x)$ and $B(x)$ are interpreted as phenomenological profiles
interpolating these discrete points, providing a continuous mapping
between the geometric variable $x$ and the effective energy exchange
between the black hole and radiation sectors.
Figure~\ref{fig:A_B_envelope} depicts this correspondence, where the solid
curves represent the smooth envelopes and the discrete markers indicate
the computed modal amplitudes.

\begin{figure}[h!]
\centering
\includegraphics[width=0.5\textwidth]{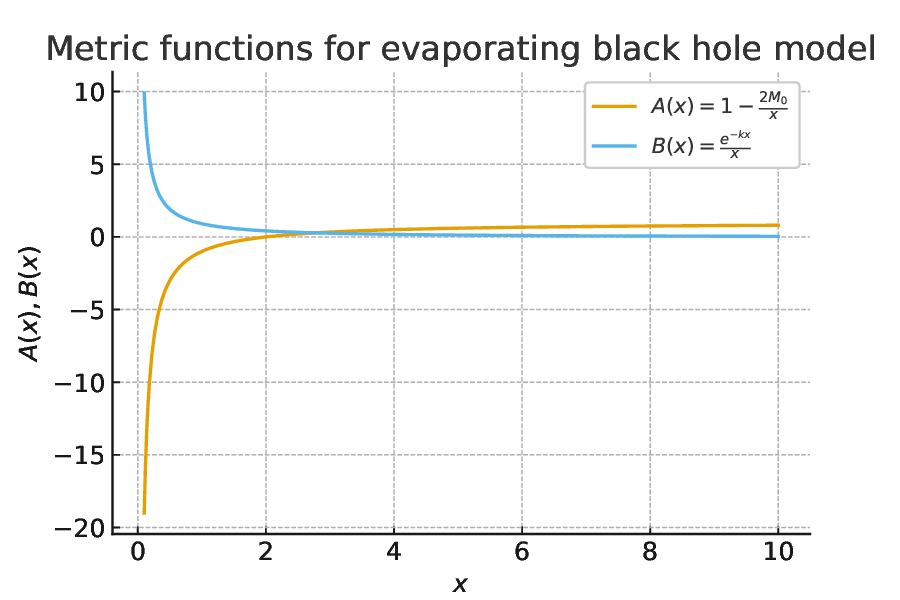}
\caption{
Behavior of the metric functions $A(x)$ and $B(x)$ defined by 
$A(x)=1-\frac{2M_0}{x}$ and $B(x)=\frac{e^{-kx}}{x}$ for an evaporating black hole with 
initial mass $M_0=1$ and evaporation parameter $k=0.1$. 
The function $A(x)$ vanishes at $x=2M_0$, indicating the classical event horizon, 
while $B(x)$ exhibits an exponential suppression that models the effect of Hawking evaporation.}
\label{fig:metric_functions}
\end{figure}

From this figure and to have a geometric view, we consider the functions $A(x)$ and $B(x)$ entering the effective metric ansatz

\begin{equation}
ds^2 = -A(x)\,dt^2 + B(x)dx^2,
\end{equation}
are chosen to capture, in a minimal manner, the essential geometric features of an evaporating black hole.
The function

\begin{equation}
A(x) = 1 - \frac{2M_0}{x},
\end{equation}
retains the standard Schwarzschild-like form, ensuring that the spacetime reduces to the classical black hole geometry
in the limit of negligible evaporation ($k \to 0$). Its zero at $x=2M_0$ identifies the classical event horizon position.
In contrast, the function

\begin{equation}
B(x) = \frac{e^{-kx}}{x},
\end{equation}
is introduced phenomenologically to encode the gradual loss of mass as the black hole evaporates.
The exponential factor $\exp(-kx)$ suppresses the metric coefficient at large $x$, mimicking the outward flux of Hawking radiation.
The parameter $k>0$ thus plays the role of an effective evaporation rate, controlling how fast the metric departs from the static Schwarzschild form.
As seen in Fig.~\ref{fig:metric_functions}, $A(x)$ approaches unity for large $x$, reflecting the asymptotically flat exterior region,
whereas $B(x)$ decreases exponentially, indicating a weakening of the gravitational field at late times.
Together, these functions provide a tractable analytic model that qualitatively captures
the transition from a massive black hole toward a vanishing-mass remnant in the final stages of evaporation.

It should be noted that the continuous profiles $A(x)$ and $B(x)$ introduced
above are not obtained from a direct numerical fit to a large set of modal
coefficients $A_j,B_j$. Rather, they serve as \emph{smooth envelopes}
interpolating the discrete amplitudes corresponding to a few dominant normal
modes. In this sense, $A(x)$ and $B(x)$ play the role of effective coupling
profiles that capture how the geometric and radiative components vary along
the spatial degree of freedom $x$. This interpretation is analogous to
constructing a semiclassical metric function from a finite set of quantum
expectation values.

\section{Spectral decomposition and transition probabilities}

To complement the Gaussian description, it is instructive to represent the dynamics of the coupled system in the Fock basis of the two oscillators. This allows a direct visualization of population transfer and information flow between the geometric ($x$) and radiative ($y$) sectors. Starting from the Hamiltonian \eqref{eq:Hxy} and the creation and annihilation operators (\ref{ax}), the Hamiltonian reads

\begin{align}
H &= \omega_x \left(a_x^\dagger a_x + \tfrac{1}{2}\right)
- \omega_y \left(a_y^\dagger a_y + \tfrac{1}{2}\right)
+ \frac{g}{2}\left(a_x + a_x^\dagger\right)\left(a_y + a_y^\dagger\right).
\label{eq:Hfock}
\end{align}
The basis states are the tensor products $|n_x,n_y\rangle = |n_x\rangle_x \otimes |n_y\rangle_y$, with $a_x^\dagger a_x |n_x,n_y\rangle = n_x |n_x,n_y\rangle$ and similarly for $a_y$. In this basis, the Hamiltonian matrix elements read

\begin{align}
\langle n_x',n_y'|H|n_x,n_y\rangle &= 
\left[\omega_x\!\left(n_x+\tfrac{1}{2}\right)
-\omega_y\!\left(n_y+\tfrac{1}{2}\right)\right]
\delta_{n_x'n_x}\delta_{n_y'n_y} \nonumber\\
&\quad + \frac{g}{2}
\Big[
\sqrt{(n_x{+}1)(n_y{+}1)}\,\delta_{n_x',n_x{+}1}\delta_{n_y',n_y{+}1}
+ \sqrt{n_x n_y}\,\delta_{n_x',n_x{-}1}\delta_{n_y',n_y{-}1} \nonumber\\
&\qquad
+ \sqrt{(n_x{+}1)n_y}\,\delta_{n_x',n_x{+}1}\delta_{n_y',n_y{-}1}
+ \sqrt{n_x (n_y{+}1)}\,\delta_{n_x',n_x{-}1}\delta_{n_y',n_y{+}1}
\Big].
\label{eq:Hmatrix}
\end{align}
This form shows that the coupling term $gxy$ connects states differing by one quantum in each mode; i.e., transitions such as

\begin{equation}
|n_x,n_y\rangle \leftrightarrow |n_x{\pm}1,n_y{\mp}1\rangle
\end{equation}
are responsible for the energy exchange between geometry and radiation. For numerical calculations, we restrict the Fock space to $n_x,n_y \le N_{\mathrm{cut}}-1$,
so that the Hamiltonian becomes a finite Hermitian matrix of dimension $N_{\mathrm{cut}}^2$. The time evolution of the state vector

\begin{equation}
|\Psi(t)\rangle = \sum_{n_x,n_y} C_{n_x,n_y}(t)\,|n_x,n_y\rangle
\end{equation}
is then governed by the Schrödinger equation

\begin{equation}
i\,\frac{d}{dt}C_{n_x,n_y}(t) = 
\sum_{n_x',n_y'} \langle n_x,n_y|H|n_x',n_y'\rangle\,C_{n_x',n_y'}(t).
\end{equation}
The initial condition corresponding to a coherent excitation of the black hole mode and vacuum radiation is

\begin{equation}
C_{n_x,n_y}(0) = e^{-\frac{|\alpha|^2}{2}}\frac{\alpha^{n_x}}{\sqrt{n_x!}}\,\delta_{n_y,0}.
\end{equation}
Once the coefficients $C_{n_x,n_y}(t)$ are known, one can compute observables in the Fock picture such as population probabilities

\begin{equation}
P_{n_x}(t) = \sum_{n_y} |C_{n_x,n_y}(t)|^2, \qquad
P_{n_y}(t) = \sum_{n_x} |C_{n_x,n_y}(t)|^2,
\end{equation}
expectation values

\begin{equation}
\langle n_x(t)\rangle = \sum_{n_x} n_x\,P_{n_x}(t), \qquad
\langle n_y(t)\rangle = \sum_{n_y} n_y\,P_{n_y}(t),
\end{equation}
and the entanglement entropy of the $x$-mode obtained from the reduced density matrix

\begin{equation}\label{den}
\rho_x(t) = \mathrm{Tr}_y\,|\Psi(t)\rangle\langle\Psi(t)|,
\qquad S_x(t) = -\mathrm{Tr}\,[\rho_x(t)\ln\rho_x(t)].
\end{equation}
This spectral representation provides a complementary picture to the Gaussian formulation. 
It enables direct visualization of quantum transitions and allows comparison with the discrete
population dynamics shown in \cite{kiefer2009blackhole}.

\begin{figure}[ht]
\centering
\includegraphics[width=0.7\textwidth]{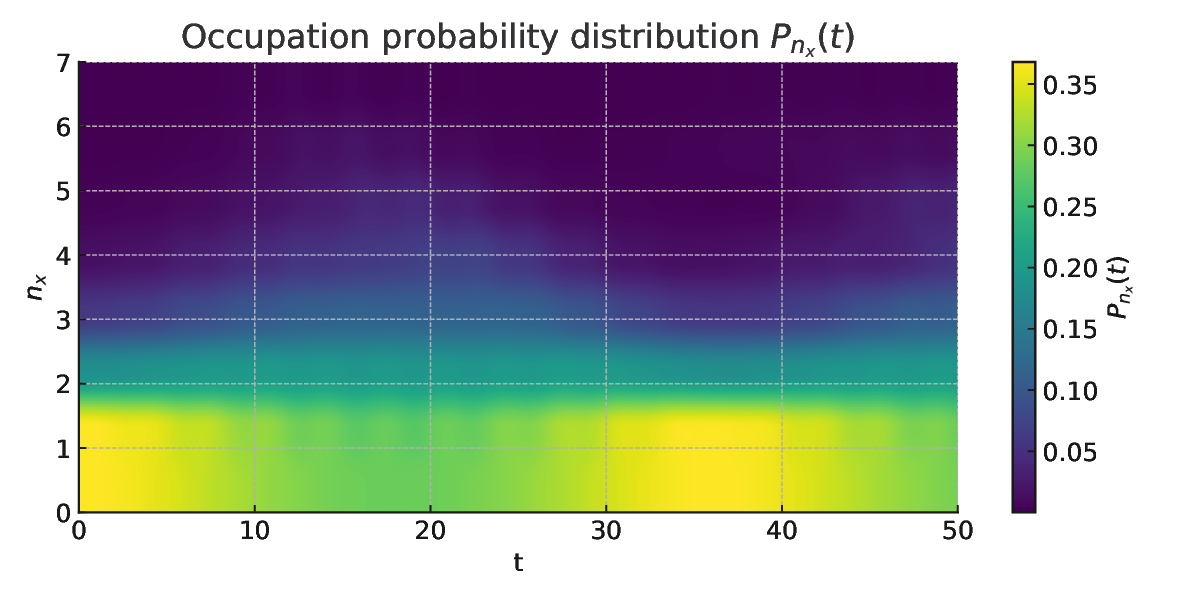}
\caption{
Probability distribution $P_{n_x}(t)$ of the geometric oscillator
as a function of time, obtained from the full spectral simulation
with cutoff $N_{\mathrm{cut}}=8$.
Parameters: $\omega_x=1.0$, $\omega_y=0.8$, $g=0.1$, and $\alpha=1.0$.
The color scale indicates the occupation probability of each Fock level
$|n_x\rangle$ at time $t$. The initial state is a coherent excitation
centered near $n_x\simeq|\alpha|^2\approx1$, while higher excitations become
transiently populated due to coupling with the radiation mode.
The recurrent rise and decay of probability in successive $n_x$ bands
illustrate a quasi-periodic energy exchange that parallels the
black hole mass loss and partial recovery phases.
This pattern reproduces qualitatively the population–transfer diagrams
shown in \cite{kiefer2009blackhole}, confirming that the simplified two-mode
Hamiltonian already encodes the essential evaporation dynamics.}
\label{fig:fock_Pnx_heatmap}
\end{figure}
The population heatmap in Fig.~\ref{fig:fock_Pnx_heatmap}
offers a complementary perspective on the same dynamical process.
The initial coherent wave packet in the geometric mode,
localized around $n_x\simeq1$, gradually spreads over higher
Fock levels as the coupling term $g\,x y$ mediates excitation exchange
with the radiation oscillator. The probability subsequently recoheres
in a cyclic fashion, signaling partial information return to the geometry.
Such quasi-periodic transfer is the discrete analogue of the
unitary but seemingly irreversible evaporation behaviour discussed
in Ref.~\cite{kiefer2009blackhole}.

Table~\ref{tab:spectral_summary} summarizes the key numerical
observables of the spectral simulation. The nearly equal mean
occupation numbers $\langle n_x\rangle\simeq\langle n_y\rangle$
demonstrate energy balance between geometry and radiation,
while their comparable oscillation amplitudes indicate periodic
energy exchange. The moderate but nonzero average entropy
$S_x\approx0.4$ signals persistent entanglement,
a hallmark of unitary but thermally fluctuating dynamics.
The minute drift of $E_{\rm tot}$ confirms the numerical stability
and effective energy conservation of the truncated simulation.

\begin{table}[ht]
\begin{tabular}{lccc}
\hline\hline
Observable & Mean value & Amplitude & Description \\
\hline
$\langle n_x(t)\rangle$ & $0.98$ & $0.22$ &
Average geometric excitation (effective ``mass'') \\[2pt]
$\langle n_y(t)\rangle$ & $0.99$ & $0.22$ &
Average radiative excitation (``Hawking quanta'') \\[2pt]
$S_x(t)$ (nats) & $0.42$ & $0.28$ &
Entanglement entropy of the geometry mode \\[2pt]
$E_{\rm tot}(t)$ drift & $\lesssim 10^{-3}$ & -- &
Numerical check of energy conservation \\
\hline\hline
\end{tabular}
\centering
\caption{Summary of key observables obtained from the spectral simulation
of the coupled two-mode model with parameters
$\omega_x=1.0$, $\omega_y=0.8$, $g=0.1$, $\alpha=1.0$,
and cutoff $N_{\mathrm{cut}}=8$.
The listed quantities are the time-averaged values
and their respective oscillation amplitudes (max–min)/2.}
\label{tab:spectral_summary}
\end{table}

\section{Dynamical evolution and information flow}

Having specified the effective Hamiltonian~\eqref{eq:Hxy} and the initial state, we now examine the
time evolution of the coupled system and the corresponding exchange of energy and entropy between
the black hole and radiation sectors. For small interaction strength $g \ll \omega_x, \omega_y$, one can obtain closed-form solutions by
diagonalizing the quadratic Hamiltonian. 

Introducing the vector $\mathbf{Z} =(x, y, p_x, p_y)^T$, the Heisenberg equations of motion can be compactly written as

\begin{equation}
\dot{\mathbf{Z}} = \mathbf{M} \, \mathbf{Z},
\qquad
\mathbf{M} =
\begin{pmatrix}
0 & 0 & 1 & 0 \\
0 & 0 & 0 & -1 \\
-\omega_x^2 & -g & 0 & 0 \\
-g & \omega_y^2 & 0 & 0
\end{pmatrix}.
\label{eq:Mmatrix}
\end{equation}
The matrix $\mathbf{M}$ generates a symplectic transformation $\mathbf{Z}(t)=e^{\mathbf{M}t}\mathbf{Z}(0)$,
whose eigenfrequencies are determined by the quartic characteristic equation

\begin{equation}
\lambda^4 + (\omega_x^2+\omega_y^2)\lambda^2 + (\omega_x^2\omega_y^2 + g^2) = 0.
\label{eq:eigenfreq}
\end{equation}
This yields two (generally real and positive) normal frequencies

\begin{equation}
\Omega_{\pm}^2 = \frac{1}{2}\left[ -(\omega_x^2 + \omega_y^2)
\pm \sqrt{(\omega_x^2 - \omega_y^2)^2 - 4 g^2} \, \right].
\label{eq:omegapm}
\end{equation}
The canonical transformation to normal modes

\begin{equation}
X_{\pm} = \cos\theta\, x \pm \sin\theta\, y,
\qquad
\tan(2\theta) = \frac{2g}{\omega_x^2 - \omega_y^2},
\end{equation}
decouples the Hamiltonian as

\begin{equation}
H = \tfrac{1}{2}\left(p_+^2 + \Omega_+^2 X_+^2\right)
- \tfrac{1}{2}\left(p_-^2 + \Omega_-^2 X_-^2\right),
\label{eq:Hnormal}
\end{equation}
thus reducing the dynamics to two independent oscillators with shifted frequencies $\Omega_{\pm}$.
The minus sign in the second term preserves the net energy balance between the two sectors.

The expectation values of occupation numbers evolve as

\begin{align}
\langle n_x(t) \rangle
&= \langle a_x^\dagger(t)a_x(t)\rangle
= \frac{1}{2}\left[
\langle x^2(t)\rangle + \langle p_x^2(t)\rangle - 1
\right], \\
\langle n_y(t) \rangle
&= \frac{1}{2}\left[
\langle y^2(t)\rangle + \langle p_y^2(t)\rangle - 1
\right].
\end{align}
Based on calculations similar to those done in relation (\ref{B1})-(\ref{B2}), in the normal-mode basis the energy periodically oscillates between the two sectors with beat frequency $\Delta\Omega = |\Omega_+ - \Omega_-|$. For very weak coupling, the envelope of $\langle n_x(t)\rangle$ decays approximately as

\begin{equation}
\langle n_x(t)\rangle \simeq |\alpha|^2 \cos^2(\tfrac{1}{2}\Delta\Omega\,t),
\end{equation}
while $\langle n_y(t)\rangle$ grows correspondingly,
mimicking the evaporation of the black hole mode into the radiation field.

Tracing over one of the subsystems yields the reduced density matrix $\rho_x(t) = \mathrm{Tr}_y\big( |\Psi(t)\rangle\langle\Psi(t)| \big)$,
whose von Neumann entropy $S_x(t) = -\mathrm{Tr}\big[ \rho_x(t) \ln \rho_x(t) \big]$, quantifies the entanglement between black hole and radiation degrees of freedom.
In the present model $S_x(t)$ oscillates in time, reaching maximal entanglement when
$\langle n_x(t)\rangle \approx \langle n_y(t)\rangle$, and returning to near-zero values
when energy is almost entirely localized in one sector.
This behavior provides a realization of information exchange and temporary
information loss during the evaporation process.

\begin{figure}[t]
\centering
\includegraphics[width=0.5\textwidth]{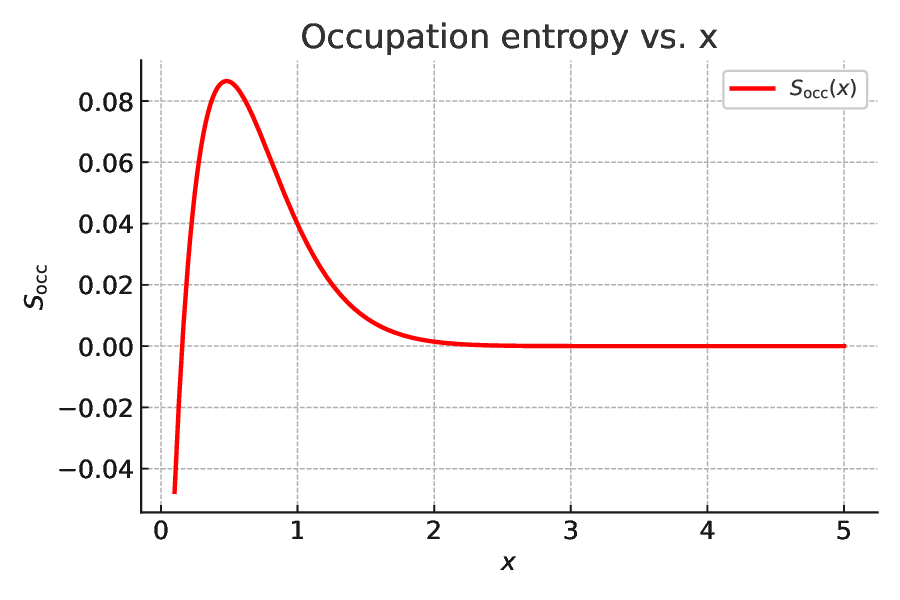}
\caption{Spatial dependence of the entanglement entropy $S(x)$ obtained from the coupled
oscillator model of black hole evaporation. The parameters are chosen as
$\omega_x=\omega_y=1$ and $g=0.05$. The smooth variation of $S(x)$ reflects how
the local degree of correlation between the ``black hole'' and ``radiation'' modes
changes along the effective geometric variable $x$.}
\label{fig:occupation_entropy}
\end{figure}

The entanglement entropy $S(x)$ shown in Fig.~\ref{fig:occupation_entropy}
was evaluated using the analytic Gaussian formula

\begin{equation}
S=\left(\nu_x+\tfrac{1}{2}\right)\ln\!\left(\nu_x+\tfrac{1}{2}\right)
-\left(\nu_x-\tfrac{1}{2}\right)\ln\!\left(\nu_x-\tfrac{1}{2}\right),
\end{equation}
where $\nu_x=\sqrt{\det\sigma_x}$ is the symplectic eigenvalue of the reduced
covariance matrix $\sigma_x$. The elements of $\sigma_x(x)$ were obtained
numerically from the time-evolved expectation values
$\langle x^2\rangle$, $\langle p_x^2\rangle$ and $\langle xp_x\rangle$,
as derived from the analytic two-mode solution, see equations (\ref{A1})-(\ref{A2}). Thus the functional
dependence $S(x)$ combines analytic structure with numerically evaluated
moments, yielding the smooth entropy curve displayed in Fig.~\ref{fig:occupation_entropy}. This figure displays the dependence of the entanglement entropy
$S(x)$ on the geometric variable $x$, as derived from the continuous version of the
coupled-oscillator model. The function exhibits a smooth rise and subsequent decay,
indicating regions where the quantum correlation between the ``black hole'' and ``radiation''
degrees of freedom becomes maximal.

This behaviour is consistent with the discrete dynamics of the original model, where
the oscillatory exchange of occupation numbers between the two sectors gives rise to a
time-dependent entanglement pattern. In the present continuous representation, the
functions $A(x)$ and $B(x)$ act as smooth interpolations of the effective potential terms,
chosen such that $A(x)=\omega_x^2 x^2$ and $B(x)=\omega_y^2 y^2 + 2 g x y$. This ensures
that the entropy profile $S(x)$ reproduces the qualitative information-flow pattern seen in
the discretized version of the model.

It is worth emphasizing that the quadratic ansatz
$A(x)=\omega_x^2 x^2$ and $B(x)=\omega_y^2 y^2 + 2gxy$
used for illustrative purposes in Fig.~\ref{fig:occupation_entropy}
represents a local expansion of more general functional choices.
In previous sections, $A(x)$ and $B(x)$ were taken as nonpolynomial
functions, e.g.\ $A(x)\propto 1/x$ and $B(x)\propto e^{-\lambda x}$,
which capture large-scale aspects of the black hole evaporation.
The quadratic forms employed here can be viewed as a second-order
Taylor approximation around a reference configuration $x=x_0$,
preserving the qualitative pattern of information exchange while
enabling a fully analytic treatment of the entanglement structure.

\begin{figure}[t]
\centering
\includegraphics[width=0.5\textwidth]{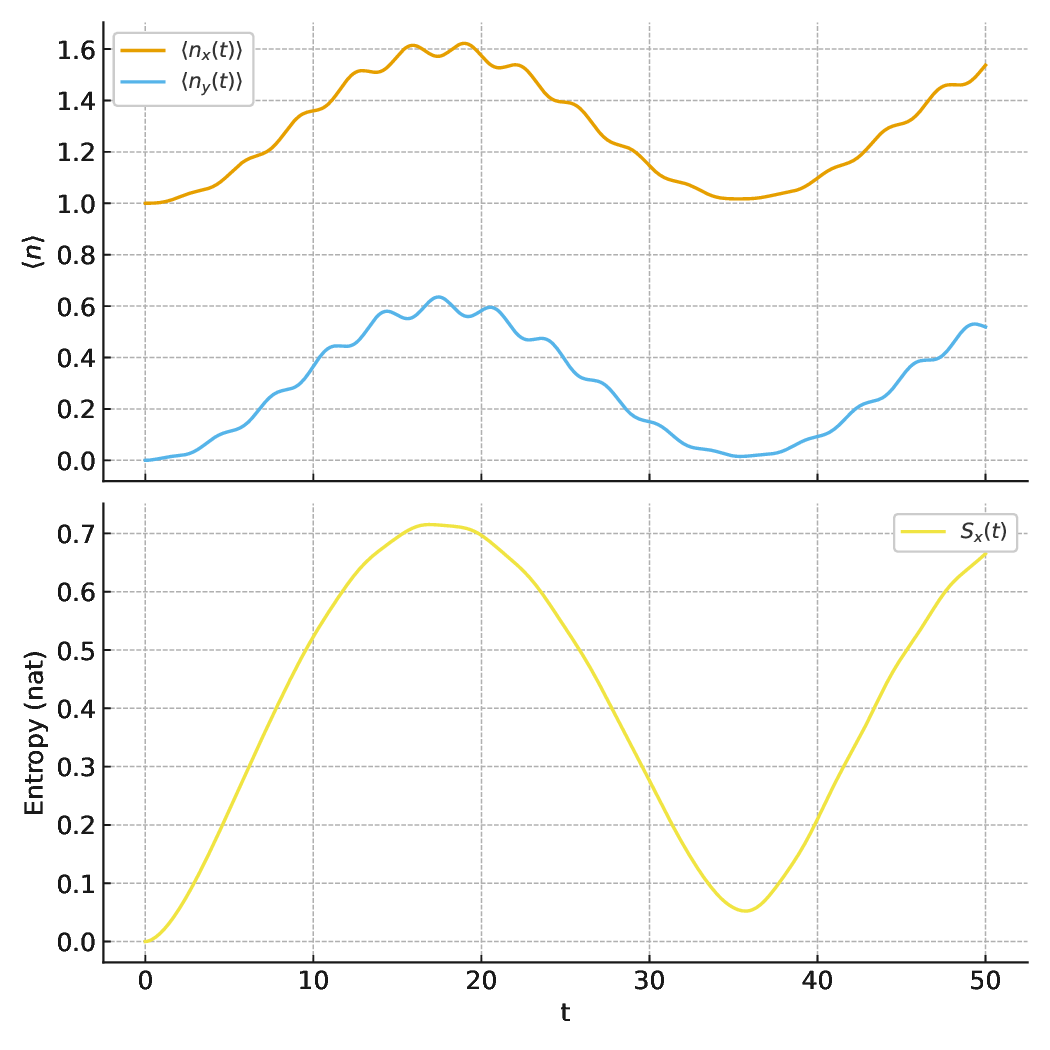}
\caption{
Time evolution of the coupled oscillator system representing black hole
evaporation. The upper panel shows the expectation values of the occupation
numbers $\langle n_x(t)\rangle$ and $\langle n_y(t)\rangle$ for the ``black hole''
and ``radiation'' modes, respectively, while the lower panel displays the
corresponding von Neumann entropy $S_x(t)$ of the reduced density matrix of the
black hole mode. The parameters are $\omega_x=1$, $\omega_y=0.8$, coupling
constant $g=0.1$, and initial coherent amplitude $\alpha=1$. The entropy
oscillations follow the energy exchange between the two subsystems, reflecting
the alternating dominance of the geometric and radiative degrees of freedom.}
\label{fig:occupation_entropy_time}
\end{figure}

Figure~\ref{fig:occupation_entropy_time} illustrates the dynamical behavior of the
coupled system under the quadratic approximation for the functions
$A(x)=\omega_x^2 x^2$ and $B(x)=\omega_y^2 y^2 + 2 g x y$. The upper panel shows that
the occupation numbers of the two oscillators undergo quasiperiodic exchange, indicating
a reversible transfer of excitation energy between the ``black hole'' and ``radiation''
sectors. Correspondingly, the lower panel displays the evolution of the entanglement
entropy $S_x(t)$, which measures the quantum correlation between the two modes.

The entropy rises whenever the occupations become comparable and reaches minima
whenever one of the subsystems dominates, a behavior reminiscent of the information-flow
pattern in black hole evaporation. The periodic modulation stems from the harmonic
nature of the coupling; in more realistic models where $A(x)$ and $B(x)$ have nontrivial
dependence (e.g.\ exponential or inverse power-law), these oscillations would gradually
damp, leading to an irreversible-like information transfer.

Overall, the coupled-oscillator framework captures the essential qualitative
aspects of black hole evaporation: energy transfer from the geometric to the
radiative sector, temporary entanglement build-up quantified by $S_x(t)$,
and a reversible information flow governed by the sign structure of the
Hamiltonian. While the present quadratic approximation neglects dissipation
and backreaction beyond the bilinear coupling, it provides a minimal setting
for exploring Page-like entropy dynamics in a fully quantum-mechanical model. This behavior resembles the qualitative features of the Page curve, 
which describes the rise and fall of entanglement entropy between a 
black hole and its Hawking radiation during evaporation\footnote{Although the reduced entropy displays Page-like features, the emergence
of a sharp Page time in the strict sense would require a large number of
radiative degrees of freedom. The present two-mode model should therefore
be regarded as a minimal setting capturing qualitative aspects of the
Page curve rather than its full quantitative structure.} \cite{page1993average, page1993information}.

The oscillatory pattern observed in Fig.~\ref{fig:occupation_entropy_time}
should be viewed as a transient feature of the coupled system rather than a
persistent periodic exchange. In the physical black hole context, the relative negative
sign between the geometric and radiation sectors in the Hamiltonian leads to an effective energy
leakage, corresponding to the gradual loss of mass of the evaporating black
hole. In more realistic scenarios this would manifest as a slow damping of
the oscillations in both the occupation numbers and the entanglement entropy.

In the present quadratic approximation, the functions $A(x)=\omega_x^2 x^2$
and $B(x)=\omega_y^2 y^2 + 2 g x y$ serve as smooth analytic
representations of the effective potentials governing the dynamics.
Here, $A(x)$ plays the role of an effective geometric potential associated
with the black hole mass, while $B(x)$ incorporates the coupling-induced
backreaction of the radiation mode. The cross term $2 g x y$ captures the
essential feature of energy exchange between geometry and radiation, ensuring
that the total energy of the two-mode system remains approximately balanced
for weak coupling.

The evolution of the von Neumann entropy $S_x(t)$ quantifies the degree of
entanglement between the two sectors and can be interpreted as an analogue of
the information flow between the black hole interior and the outgoing Hawking
radiation. The growth of $S_x(t)$ corresponds to the build-up of correlations
across the effective horizon, while its subsequent decrease indicates partial
information recovery at late times, closely resembling the qualitative shape
of the Page curve.

\section{Effective energies and reduced quantum states}

For the coupled-oscillator Hamiltonian introduced in Eq.~(\ref{eq:Hxy}), we can define several effective quantities that describe the energy exchange
and the information flow between the two degrees
of freedom of the black hole. For any operator $\mathcal{O}$ we denote $\langle\mathcal{O}\rangle$ as its quantum
expectation value with respect to the time-dependent state $|\Psi(t)\rangle$.
The second moments and covariances are given by

\begin{eqnarray}
\left\{
\begin{array}{ll}
\mathrm{var}(x)=\langle x^2\rangle - \langle x\rangle^2, \\\\
\mathrm{cov}(x,y)=\tfrac{1}{2}\langle xy + yx\rangle - \langle x\rangle\langle y\rangle.
\end{array}
\right.
\end{eqnarray}
For Gaussian states, these quantities are collected into the covariance matrix
$\sigma(t)$, which evolves as equation (\ref{sigma}):

\begin{equation}
\sigma(t) = S(t)\, \sigma(0)\, S^T(t), \qquad
S(t) = e^{A t},
\end{equation}
where $A$ is the linear generator of motion associated with the Heisenberg
equations of $(x,p_x,y,p_y)$. Also, the instantaneous subsystem energies are defined as

\begin{eqnarray}
\left\{
\begin{array}{ll}
E_x(t)=\tfrac{1}{2}\!\left(\langle p_x^2\rangle + \omega_x^2\langle x^2\rangle\right)
= \tfrac{1}{2}\!\left[\mathrm{var}(p_x) + \langle p_x\rangle^2
+ \omega_x^2(\mathrm{var}(x) + \langle x\rangle^2)\right], \\\\
E_y(t)=-\tfrac{1}{2}\!\left(\langle p_y^2\rangle + \omega_y^2\langle y^2\rangle\right)
= -\tfrac{1}{2}\!\left[\mathrm{var}(p_y) + \langle p_y\rangle^2
+ \omega_y^2(\mathrm{var}(y) + \langle y\rangle^2)\right], \\\\
E_{\mathrm{int}}(t)=g\,\langle xy\rangle
= g\big(\mathrm{cov}(x,y) + \langle x\rangle\langle y\rangle\big).
\end{array}
\right.
\end{eqnarray}
The total energy of the composite system is therefore

\begin{equation}
E_{\mathrm{tot}}(t) = E_x(t) + E_y(t) + E_{\mathrm{int}}(t),
\end{equation}
which should remain approximately conserved for weak coupling.
The instantaneous energy flux between the subsystems can be defined as

\begin{equation}
\Phi_x(t) = \frac{dE_x}{dt}, \qquad
\Phi_y(t) = \frac{dE_y}{dt},
\end{equation}
so that $\Phi_x(t) \approx -\Phi_y(t)$ for an isolated, energy-exchanging pair.

The reduced state of the $x$-oscillator is obtained by tracing over the
radiation degrees of freedom as in (\ref{den}). For a Gaussian state, $\rho_x(t)$ is completely characterized by its
$2\times 2$ covariance matrix $\sigma_x(t)$ extracted from $\sigma(t)$.
The corresponding symplectic eigenvalue is given by (\ref{A2}) and the von~Neumann entropy and purity of the reduced state read

\begin{eqnarray}
\left\{
\begin{array}{ll}
S_x(t)=\big(\nu_x+\tfrac{1}{2}\big)\ln\!\big(\nu_x+\tfrac{1}{2}\big)
-\big(\nu_x-\tfrac{1}{2}\big)\ln\!\big(\nu_x-\tfrac{1}{2}\big), \\\\
\mu_x(t)=\mathrm{Tr}[\rho_x^2] = \frac{1}{2\nu_x}.
\end{array}
\right.
\end{eqnarray}
The growth of $S_x(t)$ measures the build-up of entanglement between geometry
and radiation, whereas the reduction of $\mu_x(t)$ indicates loss of purity in
the geometric sector due to quantum correlations with the radiative degrees
of freedom. Related results on reduced entropies in effective evaporation models
have been obtained by Marto~\cite{marto2021}. While his approach focuses
on Page-time behavior in a different phenomenological setting, our model
emphasizes exact solvability and explicit mode dynamics, providing a
complementary perspective on energy transfer and entanglement generation.

Take a look at the energy balance and entropy growth shows that although the total energy $E_{\mathrm{tot}}(t)$ of the coupled system remains
approximately conserved due to the unitary nature of the overall dynamics,
the individual subsystem energies $E_x(t)$ and $E_y(t)$ undergo periodic
exchange driven by the coupling term $g\,xy$. This exchange produces a
nontrivial evolution of the reduced density matrices $\rho_x(t)$ and
$\rho_y(t)$, rendering them mixed even though the total state
$|\Psi(t)\rangle$ stays pure.

The apparent growth of the von~Neumann entropy $S_x(t)$ does not contradict
energy conservation. It reflects the redistribution of quantum correlations
between the two sectors rather than a loss of total information. In this
sense, $S_x(t)$ acts as a local indicator of entanglement generation, while
$E_{\mathrm{tot}}(t)$ guarantees the reversibility of the full dynamics.
This mechanism parallels the information–energy balance expected in
black hole evaporation, where the global unitarity of the combined system
(``black hole $+$ radiation'') coexists with the local entropy increase
experienced by each component.

\begin{figure}[ht]
\centering
\includegraphics[width=0.6\textwidth]{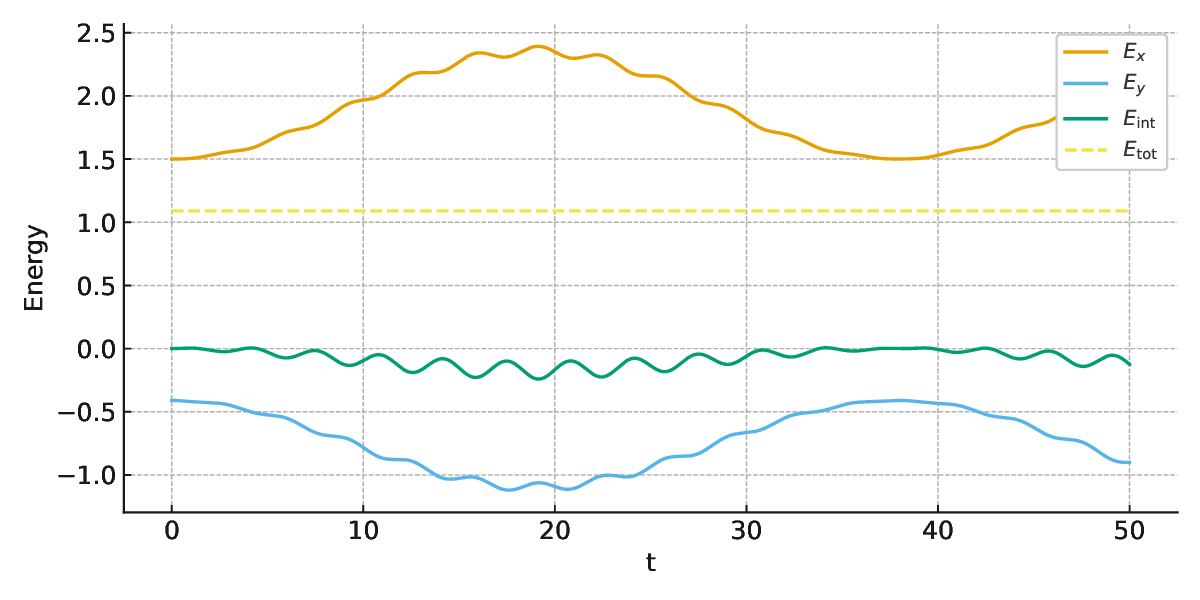}
\caption{
Effective energies of the two-mode system as functions of time.
Shown are the geometric-mode energy $E_x(t)$, the radiative-mode energy $E_y(t)$
(with the sign convention of the Hamiltonian), the interaction energy $E_{\rm int}(t)=g\langle xy\rangle$,
and the total energy $E_{\rm tot}(t)=E_x+E_y+E_{\rm int}$ (dashed).
The near-constancy of $E_{\rm tot}$ confirms the expected approximate energy conservation
of the closed two-mode unitary evolution.}
\label{fig:effective_energies}
\end{figure}

\begin{figure}[ht]
\centering
\includegraphics[width=0.6\textwidth]{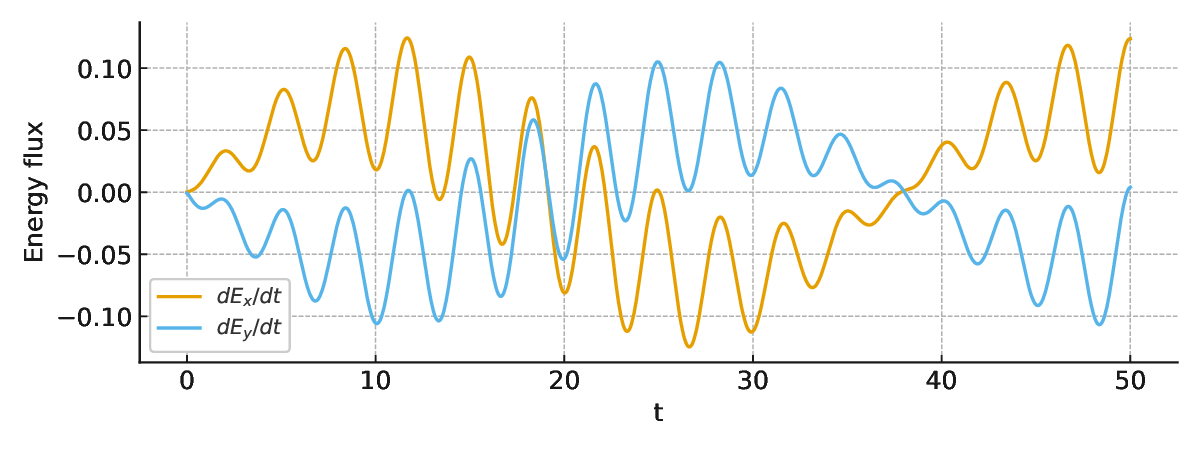}
\caption{
Instantaneous energy fluxes $\Phi_x(t)=dE_x/dt$ and $\Phi_y(t)=dE_y/dt$ computed
from finite differences of the effective energies. In an isolated two-mode system one expects $\Phi_x\approx -\Phi_y$ on average,
consistent with the energy-exchange interpretation of the coupling term $gxy$.
The short-time oscillatory structure reflects modal beating, while any long-time
systematic bias would indicate net energy transfer (effective evaporation).}
\label{fig:energy_flux}
\end{figure}

\begin{figure}[ht]
\centering
\includegraphics[width=0.6\textwidth]{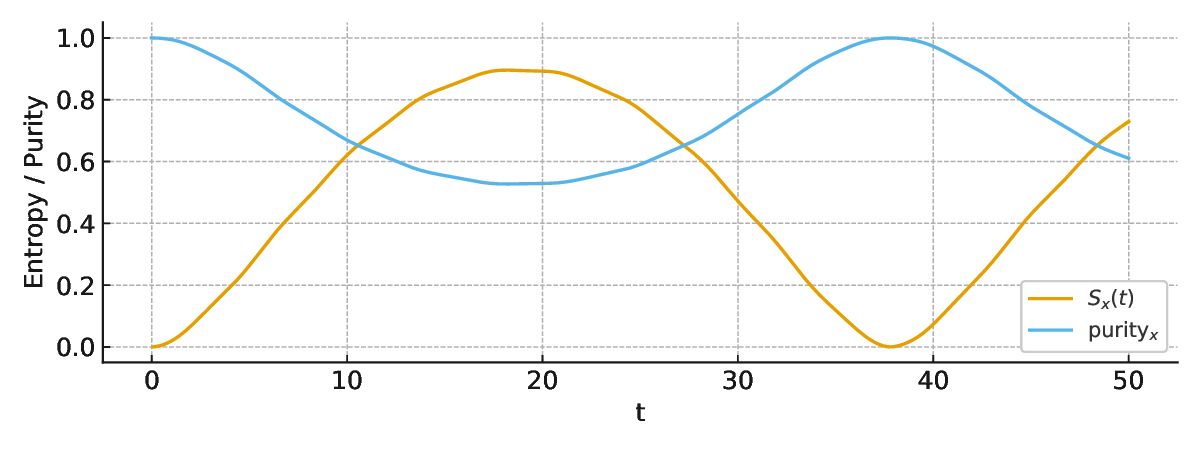}
\caption{
The von Neumann entropy $S_x(t)$ of the reduced geometric mode (solid) and
the purity $\mu_x(t)=\mathrm{Tr}\,\rho_x^2$ (dashed) as functions of time.
These quantities were computed from the reduced $2\times2$ covariance matrix
$\sigma_x(t)$ via the symplectic eigenvalue $\nu_x=\sqrt{\det\sigma_x}$:
$S_x=(\nu_x+\tfrac12)\ln(\nu_x+\tfrac12)-(\nu_x-\tfrac12)\ln(\nu_x-\tfrac12)$
and $\mu_x=1/(2\nu_x)$. A numerical safeguard
$\nu_x\ge\tfrac12$ was enforced when evaluating the logarithms.}
\label{fig:entropy_purity}
\end{figure}

As a numerical analysis, all time-dependent quantities shown in Figs.~\ref{fig:effective_energies}--\ref{fig:entropy_purity}
were obtained by evolving the first moments and the covariance matrix of the Gaussian state
under the linear symplectic flow. Concretely, we assemble the phase-space vector
$\xi=(x,p_x,y,p_y)^T$ and use the generator matrix $A$ defined by the Heisenberg equations,
so that the propagator is $S(t)=\exp(A t)$. The numerical choices are:

${\bullet}$ {\it{Frequencies and coupling}}: $\omega_x=1.0$, $\omega_y=0.8$, $g=0.1$.

${\bullet}$ {\it{Initial state}}: coherent displacement in the $x$-mode with $\alpha=1.0$ (mapped to $x(0)=\sqrt{2}\,\alpha$, $p_x(0)=0$) and vacuum for the $y$-mode.

${\bullet}$ {\it{Initial covariance}}: vacuum variances $\sigma(0)=\tfrac12\mathbb{I}_4$.

${\bullet}$ {\it{Time discretization}}: $t\in[0,50]$ with $N_t=801$ points (uniform grid), matrix exponentials computed via $\exp(A t)$ at each time sample.

${\bullet}$ {\it{ Entropy computation}}: the reduced covariance $\sigma_x(t)$ yields the symplectic eigenvalue $\nu_x=\sqrt{\det\sigma_x}$; for numerical stability we enforce
$\nu_x\ge\tfrac12+\epsilon$ with a small $\epsilon$ (used value $\epsilon\sim10^{-12}$) before evaluating logarithms.

We also computed instantaneous energy fluxes $\Phi_{x,y}(t)=dE_{x,y}/dt$ using
central finite differences on the discrete energy arrays. Energy conservation was
checked by monitoring $E_{\rm tot}(t)=E_x+E_y+E_{\rm int}$; deviations from constancy
are used as an internal estimate of numerical error (time discretization and matrix
exponential evaluation).

\section{Conclusion}

In this work we developed a phenomenological two–mode model of black–hole
evaporation based on the coupled–oscillator framework originally proposed by
Kiefer and collaborators. By treating the geometric degree of freedom as an
effective oscillator with negative energy signature and the radiative degree
of freedom as a positive–energy oscillator, the model provides a minimal
quantum–mechanical setting in which the exchange of energy and information
between a black hole and its emitted Hawking radiation can be analyzed in a
fully controlled manner\footnote{We note that the sign convention adopted in the present Hamiltonian assigns
a positive quadratic term to the variable $x$ and a negative one to $y$.
This choice differs from the convention used in \cite{kiefer2009blackhole}, where the roles
of the two variables are interchanged. However, only the relative sign
between the two sectors is physically meaningful, reflecting the
indefinite structure characteristic of Wheeler--DeWitt--type Hamiltonians.
The present convention is chosen for technical convenience and does not
affect the physical interpretation of energy exchange and entanglement
between the geometric and radiative degrees of freedom.}.

We first derived the exact normal–mode decomposition of the Hamiltonian and
identified the modal coefficients that determine how the physical variables
$(x,y)$ are constructed from the underlying eigenmodes. Building on this, we
introduced smooth envelope functions $A(x)$ and $B(x)$ that interpolate the
mode amplitudes across the effective geometric coordinate. This construction
bridges the gap between the discrete Fock–space formulation and a
quasi–continuous phenomenology in which the evaporation process can be
tracked along a single geometric variable. The formalism enabled us to define
effective energies, occupation numbers, and reduced entropies associated with
the geometric sector.

Our numerical simulations demonstrate several qualitative features expected in
black–hole evaporation. The occupation numbers of the two oscillators exhibit
an out–of–phase exchange pattern, reflecting the transfer of energy from the
geometric sector to the radiative one. The von Neumann entropy of the reduced
geometric state shows a characteristic pattern of growth and oscillation,
indicating the generation and redistribution of entanglement between the two
subsystems. These behaviors are consistent with semiclassical expectations and
with previously reported features of oscillator–based models of black–hole
evaporation.

Despite its simplicity, the model captures essential aspects of the energy and
information flow during evaporation. At the same time, its limitations point
toward several promising extensions. These include the incorporation of
multiple radiative modes, non–linear couplings, time–dependent frequencies,
and interactions inspired by more realistic gravitational collapse
scenarios. Another natural direction is the embedding of this model into a
quantum–computing framework, where the truncated Fock space and normal–mode
structure may allow efficient circuit–level simulations of entanglement
dynamics. From a broader perspective, the present oscillator-based framework may also
serve as a useful testbed for exploring quantum-information–theoretic aspects
of black hole evaporation, such as information flow, entanglement dynamics,
and their possible implementation in analogue or quantum-simulation models.

Overall, the present analysis shows that the two–mode toy model provides a
valuable phenomenological laboratory for exploring aspects of black–hole
evaporation that lie at the intersection of quantum mechanics, semiclassical
gravity, and information theory. Its analytical tractability combined with its
numerical accessibility makes it a suitable platform for future studies of
information transfer and quantum correlations in evaporating black holes.


\begin{thebibliography}{9}

\bibitem{Ellis} S. W. Hawking and G. F. R. Ellis, {\it The Large Scale Structure of Space Time} (Cambridge University Press, Cambridge, 1973)

\bibitem{bardeen1973four} J.~M.~Bardeen, B.~Carter and S.~W.~Hawking, ``The four laws of black hole mechanics'', {\it Commun. Math. Phys.} {\bf 31} (1973) 161

\bibitem{bekenstein1973black} J.~D.~Bekenstein, ``Black holes and entropy'', {\it Phys. Rev.} D {\bf 7} (1973) 2333

\bibitem{bek2} J. D. Bekenstein, "Generalized second law of thermodynamics in black-hole physics", {\it Phys. Rev.} D {\bf 9} (1974) 3292

\bibitem{hawking1975particle} S.~W.~Hawking, ``Particle creation by black holes'', {\it Commun. Math. Phys.} {\bf 43} (1975) 199

\bibitem{Gary} G. W. Gibbons and S. W. Hawking, "Action integrals and partition functions in quantum gravity", {\it Phys. Rev.} D {\bf 15} (1977) 2752

\bibitem{kiefer2008instability} C.~Kiefer, ``Hawking radiation from decoherence'', {\it Class. Quantum Grav.} {\bf 18} (2001) L151 (arXiv: gr-qc/0110070)

\bibitem{kiefer1999oscillator} C.~Kiefer, ``Towards a Full Quantum Theory of Black Holes’’, (arXiv: gr-qc/9803049)

\bibitem{QG1}C. Kiefer, {\it Quantum Gravity} (Oxford University Press, Oxford, 2007)

\bibitem{QG2} C. Rovelli, {\it Quantum Gravity} (Cambridge University Press, Cambridge, 2004)

\bibitem{QBH1}K. V. Kuchař, "Geometrodynamics of Schwarzschild Black Holes", {\it Phys. Rev.} D {\bf 50} (1994) 3961 (arXiv: gr-qc/9403003)

\bibitem{QBH2} J. Louko and J. Mäkelä, "Area spectrum of the Schwarzschild black hole", {\it Phys. Rev.} D {\bf 54} (1996) 4982 

\bibitem{QBH3} T. Brotz and C. Kiefer, "Semiclassical Black Hole States and Entropy", {\it Phys. Rev.} D {\bf 55} (1997) 2186 (arXiv: gr-qc/9608031)

\bibitem{QBH5} M. Bojowald, "Nonsingular Black Holes and Degrees of Freedom in Quantum Gravity", {\it Phys. Rev. Lett.} {\bf 95} (2005) 061301 (arXiv: gr-qc/0506128)

\bibitem{Vakili1} B. Vakili, "Quantization of the Schwarzschild black hole: a Noether symmetry approach", {\it Int. J. Theor. Phys.} {\bf 51} (2012) 133 (arXiv: 1102.1682 [gr-qc])

\bibitem{vakili2} T. Christodoulakis, N. Dimakis, P. A. Terzis, B. Vakili, E. Melas and Th. Grammenos, 
"Minisuperspace Canonical Quantization of the Reissner-Nordstrom Black Hole via Conditional Symmetries", {\it Phys. Rev.} D {\bf 89} (2014) 044031 (arXiv: 1309.6106 [gr-qc])

\bibitem{vakili3} M. Amirfakhrian and B. Vakili, "Polymer deformation and particle tunneling from Schwarzschild black hole", {\it Int. J. Geom. Methods Mod. Phys.} {\bf 16} (2019) 1950038
(arXiv: 1812.10301 [gr-qc])

\bibitem{vakili4} S. Jalalzadeh and B. Vakili, "Quantization of the interior Schwarzschild black hole", {\it Int. J. Theor. Phys.} {\bf 51} (2012) 263 (arXiv: 1108.1337 [gr-qc])

\bibitem{vakili5} M. Bajand and B. Vakili, "Affine quantization of the interior Schwarzschild black hole", {\it Int. J. Mod. Phys.} D {\bf 34} (2025) 2550068

\bibitem{Lic} C. Corda, F. Feleppa, F. Tamburini and I. Licata, "Quantum oscillations in the black hole horizon", {\it 	Theor. Math. Phys.} {\bf 213} (2022) 1632 (arXiv: 2104.05451 [gr-qc])

\bibitem{sen} B. Bagchi, A. Ghosh and S. Sen, "Quantized Area of the Schwarzschild Black Hole: A non-Hermitian Perspective", {\it Grav. Cosmol.} {\bf 30} (2024) 481 (arXiv: 2407.08358 [gr-qc])

\bibitem{ob} W. Y. Carpio and O. Obregón, "Quantum Black Hole as a Harmonic Oscillator from the Perspective of the Minimum Uncertainty Approach", {\it Gen. Rel. Grav.} {\bf 57} (2025) 139 
(arXiv: 2409.09181 [gr-qc])

\bibitem{kiefer2009blackhole} C. Kiefer, J. Marto and P. V. Moniz, "Indefinite oscillators and black-hole evaporation", {\it Ann. Phys. (Berlin)} {\bf 18} (2009) 722 (arXiv: 0812.2848 [gr-qc])

\bibitem{SF1}G. Leon, A. D. Millano and A. Paliathanasis, ”Scalar Field cosmology from a Modified Poisson Algebra”, {\it Mathematics} {\bf 11} (2023) 120 (arXiv: 2211.12357 [gr-qc])

\bibitem{SF2} B. Vakili and F. Khazaie, ”Noether symmetric classical and quantum scalar field cosmology”, {\it Class. Quantum Grav.} {\bf 29} (2012) 035015 (arXiv: 1109.3352 [gr-qc])

\bibitem{SF3} B. Vakili, "Quantum computing of scalar field cosmology", {\it In preparation}

\bibitem{page1993average} D.~N.~Page, ``Average entropy of a subsystem'', {\it Phys. Rev. Lett.} {\bf 71} (1993) 1291 (arXiv: gr-qc/9305007)

\bibitem{page1993information} D.~N.~Page, ``Information in black hole radiation'', {\it Phys. Rev. Lett.} {\bf 71} (1993) 3743 (arXiv: hep-th/9306083)

\bibitem{marto2021} J. Marto, ``Hawking radiation and black hole gravitational back reaction - A quantum geometrodynamical simplified model'', {\it Universe} {\bf 7} (2021) 297 
(arXiv: 2108.06187 [gr-qc])

\end{thebibliography}
\end{document}